\documentclass[aps,twocolumn,superscriptaddress]{revtex4}
\usepackage{epsfig}
\usepackage{times}
\usepackage{float}
\begin{document}

\title{Stability of cooperation under image scoring in group interactions}

\author{Heinrich H. Nax}
\email{hnax@ethz.ch}
\affiliation{Department of Social Sciences, ETH Z\"{u}rich, 8092 Z\"{u}rich, Switzerland}

\author{Matja{\v z} Perc}
\affiliation{Department of Physics, Faculty of Sciences, King Abdulaziz University, Jeddah, Saudi Arabia}
\affiliation{Faculty of Natural Sciences and Mathematics, University of Maribor, Koro{\v s}ka cesta 160, SI-2000 Maribor, Slovenia}

\author{Attila Szolnoki}
\affiliation{Institute of Technical Physics and Materials Science, Centre for Energy Research, Hungarian Academy of Sciences, P.O. Box 49, H-1525 Budapest, Hungary}

\author{Dirk Helbing}
\affiliation{Computational Social Science, ETH Z\"{u}rich, 8092 Z\"{u}rich, Switzerland}
\affiliation{Risk Center, ETH Z\"{u}rich, 8092 Z\"{u}rich, Switzerland}

\begin{abstract}
Image scoring sustains cooperation in the repeated two-player prisoner's dilemma through indirect reciprocity, even though defection is the uniquely dominant selfish behaviour in the one-shot game. Many real-world dilemma situations, however, firstly, take place in groups and, secondly, lack the necessary transparency to inform subjects reliably of others' individual past actions. Instead, there is revelation of information regarding groups, which allows for `group scoring' but not for image scoring. Here, we study how sensitive the positive results related to image scoring are to information based on group scoring. We combine analytic results and computer simulations to specify the conditions for the emergence of cooperation. We show that under pure group scoring, that is, under the complete absence of image-scoring information, cooperation is unsustainable. Away from this extreme case, however, the necessary degree of image scoring relative to group scoring depends on the population size and is generally very small. We thus conclude that the positive results based on image scoring apply to a much broader range of informational settings that are relevant in the real world than previously assumed.
\end{abstract}


\maketitle

Public goods provision \cite{Isa85,Isa88}, common-pool resource management \cite{Ost90}, and other social dilemma situations often require large groups of individuals to make individually costly contributions toward a collective action, i.e., to cooperate \cite{Olson_65}. Without sufficient consideration of future consequences, however, many real-world interactions lack the necessary cooperativeness to prevent misalignment of private and social interests. This
leads to socially undesirable outcomes. The result may be irrevocable mismanagement and over-exhaustion of shared resources, ultimately resulting in the `tragedy of the commons' \cite{hardin_g_s68, Ost90, ostrom_arps99}, which is a long-term outcome that is worse for everyone.

Hence, the `puzzle of cooperation' \cite{axelrod_84, hammerstein_03, Henrich_06, rand_tcs13, nowak_11}. Why do certain social interactions flourish with high levels of foresight and cooperation, while others are impeded by short-sighted free-riding behaviour? Game theory \cite{neumann_ma28, neumann_44} provides the necessary theoretical framework for studying the individual-level motivations \cite{nash_am51} to answer this question. Indeed, game theory provides many answers, including a variety of evolutionary \cite{Now06, west2011sixteen} and psychological \cite{fehr_n02, fehr_n03b, bowles_11} explanations, but only two based, in the language of game theory, on `rationality'.

One of these two rational explanations is based on foresightedness. Namely, if individuals take future consequences sufficiently into account, then equilibria exist supported by more sophisticated, repeated-game strategies that overcome the short-term incentives to free-ride \cite{Friedman_71}. Unfortunately, to uphold cooperation that way, individuals would need to commit to strategies in ways that may produce grim consequences if not everyone follows the strategy equilibrium path \cite{Rubinstein_79}. Hence, groups consisting of sufficiently foresighted agents may succeed to guarantee provision of a common, while groups with too many short-sighted or boundedly rational agents will fail. The second rational explanation of the puzzle of cooperation is based on building a reputation from the past \cite{Kre82}. Essentially, players build a commitment-to-cooperation reputation through their past actions, and find each other, thus, outperforming defectors (who are stuck with other defectors), even though defection is the uniquely dominant selfish behaviour in the one-shot game \cite{nash_am51}. As a result, if reputation matters sufficiently in determining with whom individuals interact, then cooperation can survive despite limited foresight.

Unfortunately, due to the inherent simultaneity of interactions \cite{And88}, it is impossible to condition one's own decision on the decisions of the others \cite{nash_am51}. One of the most important, and perhaps the simplest, reputation mechanism known in the literature to overcome this problem is image scoring \cite{nowak_n98}. Famously, image scoring can sustain cooperation in the repeated two-player prisoner's dilemma through various forms of indirect reciprocity \cite{trivers_qrb71, sugden1986, alexander_87}. Under image scoring, agents learn who cooperated and who defected in previous interactions, and consequently condition their own actions on this information. Essentially, image scoring enables cooperators to find each other, and this overcomes the negative Nash equilibrium prediction of universal defection from the one-shot game. Interestingly, image scoring has been shown to work in the laboratory \cite{wedekind_s00, milinski_prslb01, bolton2005cooperation, seinen2006social, engelmann_geb09}, but in general, it is considered to provide a relatively frail support to cooperative behaviour \cite{nowak_jtb98, panchanathan2003tale}.

In fact, many real-world social dilemmas unfold in groups \cite{perc_jrsi13}, and it is unlikely that individuals will have access to others' individual action histories \cite{Isa85, Isa88}. Information that should be readily available, however, concerns the performance of the groups as a whole. Such information thus enables `group scoring' as an alternative to image scoring. In particular, the image of an individual is no longer determined by its own past action, but by the performance of the group where an individual is member. More precisely, each player's group score summarizes the aggregate cooperativeness of the groups where he was a member in the past, without any additional information regarding what the player did individually. Two important and previously unaddressed new questions emerge: (i) How do results related to image scoring generalize to group scoring?, and (ii) How sensitive are these results to information, that is, when image scoring constitutes a proportion $p\in[0,1]$ of information made available and the residual information is based on group scoring?

The common feature of previous research on image scoring is that, over time, cooperators achieve higher scores and defectors achieve lower scores, and interactions are matched based on these scores such that thus cooperators play with cooperators and defectors play with defectors. In our paper, we build on this common feature by assuming the existence of a mechanism that assorts and matches players by their scores. In addition, we extent the scope of image-scoring-based models by analysing the sensitivity of the results to the imperfections of scores that ought to reflect individuals' true past cooperativeness rather than the overall performance of the groups where they are members. Our analysis continuously spans the worlds of two extremes; image scoring and group scoring. As we will show, image scoring, reflecting accurately individuals' past actions, works perfectly also in the generalized prisoner's dilemma game that is governed by group interaction. Conversely, group scoring fails, as it enables defectors to effectively hide behind the cooperative efforts of others in the group. But how many true images are needed for cooperation to evolve in group interaction? In other words, what is the necessary proportion $p\in(0,1)$ of image scoring? It turns out that this depends sensitively on the underlying parameters of the interactions in ways that provide a formal basis for some of Ostrom's conditions for successful common-pool resource management \cite{Ost90}. Key determinants are the rate of return, the size of the population, and the group size. Remarkably, for large populations only a `grain' of image scoring is generally sufficient for cooperation to become dominant.

\section*{Results}
We shall now formalize our arguments, generalizing step by step the two-by-two prisoner's dilemma model due to \cite{nowak_n98} to the more general context of the voluntary contributions game \cite{Isa85,Isa88}.
Suppose a population $N=\{1,2,...,n\}$ plays the following game in rounds $t=\{1,2,...,\infty\}$. Each player $i\in N$, in period $t$, chooses a contribution $c^t_i$ from budget $B=\{0,1\}$. At the same time, each player $i$ is associated with a binary \emph{score} $s_i^t=\{0,1\}$, and players are matched into $k$ groups $\mathcal{S}=\{S_1,...,S_k\}$ of a fixed size $s=n/k$ according to the ranking of players' scores (with random tie-breaking).
After groups have formed, $i$'s resulting payoff turns out $\phi^t_i(c)=(1-c^t_i )+r/s* \sum_{j\in S_i^t} c^t_j$, where $S^t_i$ is $i$'s group.
$r$ is the game's fixed rate of return, and $r/s$ the game's `marginal per-capita rate of return' that summarizes the underlying game's synergy. We assume, as is standard, $r\in [1,s]$, that is, contributing a unit of budget is socially beneficial (yielding a sum total of payoffs to all players larger than one), but individually costly.

We consider the following range of scoring mechanisms between image scoring and group scoring.

\begin{description}
\item[Image scoring:] First, we formulate the equivalent of \emph{image scoring} \cite{nowak_n98} in our setup:
at time $t$ each player $i$ has an image score, $s_{Ii}^{t}$, known to every player which is based on decisions prior to $t$.
 In period $t+1$, if $i$'s period-$t$ contribution $c_i^t$ exceeded the average contribution $\overline{c}^t=\sum_{i\in N}c_i^t/n $ at time $t$, then his image score is one; if $c_i^t<\overline{c}^t$ then it is zero. Finally, if $c_i^t=\overline{c}^t $, then $s_i^{t+1}=s_i^{t}$.

\item[Group scoring:] Analogously, we formulate \emph{group scoring}:
at time $t$ each player $i$ has a group score, $s_{Gi}^{t}$.
Suppose $i$'s period-$t$ group is $S$.
If the contribution in $S$, $c_S^t=\sum_{i\in S} c_i^t$, exceeded the average group contribution overall, $\overline{c}_S^t=\sum_{S\in \mathcal{S}^t}c_S^t/k $, then $i$'s period-$(t+1)$ group score is one; if $c_S^t<\overline{c}_S^t$ then it is zero. Finally, if $c_S^t=\overline{c}_S^t$, then $s_i^{t+1}=s_i^{t}$.

\item[Hybrid scoring:]
A hybrid between \emph{image scoring} and \emph{group scoring} in our setup means that,
at time $t$, each player $i$ has a hybrid score, $s_{Hi}^{t}$, known to every player which is based on decisions prior to $t$.
In period $t+1$, if $i$'s score is updated according to image scoring with probability $p$, and according to group scoring with probability $1-p$.
\end{description}

To summarize the three scoring methods, the types of information necessary under the different regimes are as follows. For image scoring, ex post individual-level information about contribution decisions is necessary, group-level information is therefore trivially also available. For group scoring, group associations and ex post group-level contributions need be known, individual-level information is not necessary. For the hybrid case, characterized by degree of image scoring $p\in [0,1]$, the probability that individual-level information rather than only group-level information becomes available must be larger than zero.

The stability of cooperation under the different scoring rules can be evaluated. One result is that tragedy of the commons (resulting from universal defection) is a potential risk in all cases. The stability of universal defection under all scoring rules derives from the fact that unilateral defection is a best response under all scoring methods against a state of universal defection. However, the relative stability of this worst-case outcome vis-a-vis a highly cooperative state critically depends on the scoring rule, mitigating this issue.

Under image scoring, high levels of cooperation can be stabilized and then turn out to be more stable. This is the case if the proportion of cooperators with score one (matched in good groups) grows exactly at the speed so as to neutralize the shrinking of the proportion of cooperators with score zero (matched in bad groups). The defectors profit from the contributions of the latter group and achieve an average growth equal to that of the average cooperator.

\begin{figure}
\centerline{\epsfig{file=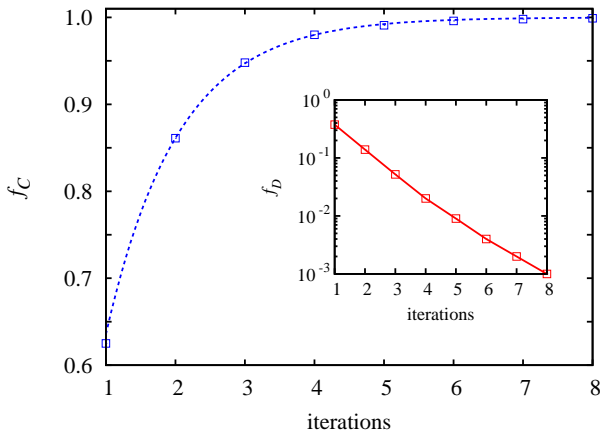,width=8.5cm}}
\caption{Under image scoring cooperators always rise to complete dominance. Shown is the fraction of cooperators in dependence on the number of iterations. The inset shows the corresponding fraction of defectors on a semi-logarithmic scale to highlight their exponential decay. Depicted results are averaged over $200$ independent realizations to minimize fluctuations. Parameter values are: $s=5$, $n=1500$ and $r=4$.}
\label{image}
\end{figure}

Stability is summarized with the results presented in Fig.~\ref{image} from simulations. The Methods section contains analytical proof of these results, as well as a description of the employed Monte Carlo simulation procedure. It can be observed that the state of complete cooperation is reached exponentially fast. We emphasize that this result is recovered independently of the value of $s$, $r$ and $n$, and it is also robust against variations of the strategy adoption rule. Cooperation will always prevail under image scoring, as it allows cooperators to separate from defectors. In general, cooperators form homogeneous groups that provide them with a competitive payoff. Conversely, defectors must be content to form groups with their like, which provides them a null payoff. Cooperators can therefore easily invade defectors, and they do so with a speed that is proportional to their number, which ultimately gives rise to the exponentially fast downfall of defectors.

At first sight, such a state of cooperation may also seem a candidate for stability under group scoring. Inspection of the individual growth dynamics, however, reveals one crucial difference.
Namely, cooperation states are not robust against the influx of defectors with score one. These players outperform all others, which, jointly with the fact that score-zero defectors outperform score-zero cooperators, implies an above-average growth rate for defection vis-a-vis cooperation. In other words, the key difference between image scoring and group scoring is that defectors can only free-ride on the contributions of others under image scoring, while, under group scoring, defectors can free-ride on the contributions \emph{and} scores of others.

\begin{figure}
\centerline{\epsfig{file=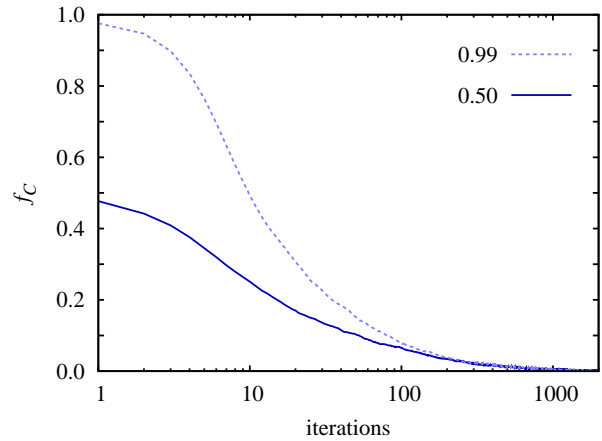,width=8.5cm}}
\caption{Under group scoring cooperators always die out. Shown is the fraction of cooperators in dependence on the number of iterations for two different initial fractions (see legend). Depicted results are averaged over $200$ independent realizations to minimize fluctuations. Parameter values are the same as in Fig.~\ref{image}.}
\label{group}
\end{figure}

Figure~\ref{group} mirrors the set-up of Fig.~\ref{image} in terms of the simulation procedure, only that here group scoring instead of image scoring is used. It can be observed that, irrespective of the initial fraction of cooperators, they eventually die out. As by image scoring, this outcome too is robust against variations of $s$, $r$, $n$ and the strategy adoption rule. Group scoring allows defectors to have the same high score as cooperators, which in turn disables the separation of the two strategies into homogeneous groups. In agreement with the outcome of the public goods game in a well-mixed population, even a single defector can therefore eventually invade the entire population. Groups scoring thus completely fails to mitigate the tragedy of the commons.

Since image scoring and group scoring could not be more different in their ability to stabilize cooperation, it remains of interest to determine the merit of hybrid scoring. While it seems reasonable to assume that sometimes the information about the past of each particular individual is readily available, more often that not the scoring of an individual is possible only indirectly through the achievements of the groups where s/he was member. We note that individual contributions in group efforts are notoriously difficult to pinpoint, which is also why the reciprocation to such efforts is quite a vague concept -- if a group contains a cooperator and a defector, who do you reciprocate with \cite{sigmund_tee07}? The question thus is, just how much individual-level information is needed to stabilize cooperation? To answer this question, we introduce the probability $p$ that a player's score is determined by image scoring, while otherwise, with probability $1-p$, group scoring is used. All other simulation details remain the same as in Figs.~\ref{image} and \ref{group}.

\begin{figure}
\centerline{\epsfig{file=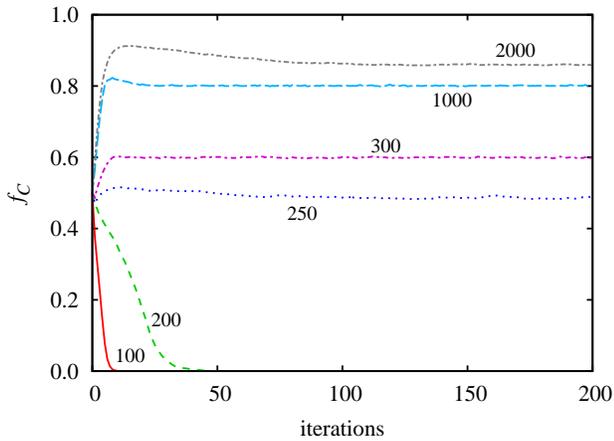,width=8.5cm}}
\caption{Under hybrid scoring, even a minute probability for accessing the image score of an individual player suffices to stabilize cooperation. However, the size of the population is crucial, as the segregation of cooperators from defectors takes the longer the smaller the value of $p$. Thus, if the population size is too small, cooperators are likely to die out before segregating into homogeneous groups. Segregation is also decelerated by low values $r$ (see Fig.~\ref{pr} for details). Shown is the fraction of cooperators in dependence on the number of iterations, as obtained for different numbers of groups forming the population (indicated alongside the lines). Depicted results are averaged over $50-200$ independent realizations to minimize fluctuations. Parameter values are: $p=0.01$, $s=5$ and $r=1.1$.}
\label{hybrid}
\end{figure}

\begin{figure}
\centerline{\epsfig{file=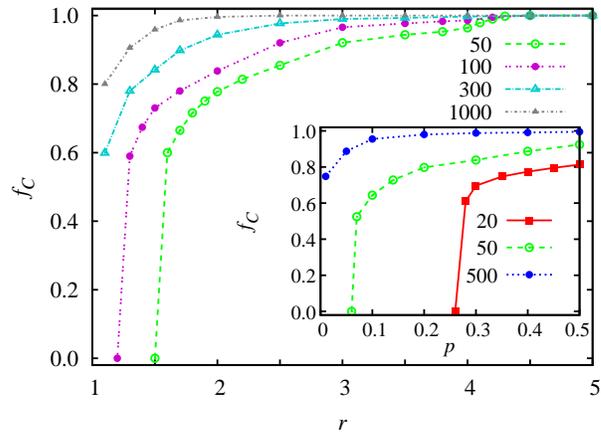,width=8.5cm}}
\caption{Hybrid scoring requires a critical population size to stabilize cooperation under adverse conditions. Cooperators essentially compete to segregate from defectors before being completely wiped out. The lower the value of $r$ (main panel) and $p$ (inset), the larger the population size needed for cooperation to be stabilized. Note that for a sufficiently small population size (see legend), there exist sudden drops to zero cooperation levels at critical values of $r$ and $p$. Shown is the stationary fraction of cooperators in dependence on $r$ (main) and $p$ (inset), as obtained for different numbers of groups forming the population (see legend). Parameter values are: $p=0.01$ (main) and $r=1.1$ (inset).}
\label{pr}
\end{figure}

Results presented in Fig.~\ref{hybrid} show that cooperation can evolve even at a very small $p$ value, if only the population size is sufficiently large. The key for the stability of cooperation is for cooperators being able to recognize each other through their high scores, and thus to form homogeneous groups. The lower the value of $p$ and the lower the value of $r$, the longer it takes for cooperators to segregate from defectors. Since cooperators are threatened by extinction, it is imperative that the segregation occurs before defectors take over. Accordingly, the lower the value of $p$ and $r$, the larger the population size needs to be to warrant sufficient time to cooperators to segregate before they die out. A lower bound for $p$ is $p\geq 2/n$. Results presented in Fig.~\ref{pr} make these arguments quantitatively more accurate. Evidently, the lower the population size, the higher the values of $p$ and $r$ need to be for cooperation to prevail. In small populations, there exist critical threshold values for both $r$ (main panel) and $r$ (inset), where drops to defector dominance are abrupt and occurring without precursors.

We emphasize that the results concerning hybrid scoring mechanisms are independent of the group size as long as their number, and hence the population size, is sufficiently large, and they are also independent of the strategy adoption rule. This corroborates our main argument, which is that, regardless of the scoring that is used, conditions need to be given for cooperators to completely segregate from defectors, i.e., to form homogeneous groups without a single defector. The identification of defectors has second-order importance. The key goal of scoring is thus to allow cooperators to recognize each other efficiently and to form homogenous groups accordingly.

\section*{Discussion}
``The most important unanswered question in evolutionary biology, and more generally in the social sciences, is how cooperative behaviour evolved and can be maintained in human or other animal groups and societies'' (Robert May in his Presidential Address to the Royal Society in 2005). A seminal explanation for the puzzle of cooperation is based on image scoring \cite{nowak_n98}, a mechanism that is both stunningly successful and stunningly simple. However, in its original formulation and application \cite{nowak_n98}, it came with the restriction to interactions that are pairwise to informational environments that allow a complete tracking of individual-level information.

In the real world, these restrictive assumptions may not be germane. Instead, cooperation may involve several groups of individuals, and ex post information regarding individual-level cooperativeness may only percolate imperfectly through group-level information. Our focus in this paper has been to determine robust theoretical predictions regarding the emergence and survival of cooperation in such situations. The presented results have rather important implications. One is negative. Namely, when individual-level information is not available, cooperation cannot spread. But there is a silver lining. When there is at least a `grain' of individual-level information, this may suffice for cooperators to find each other and form groups that are impervious to an invasion by defectors. We have shown that the spread of cooperation is robust to extensive imperfections in image scoring, thus extending the domain of environments where we should expect flourishing cooperation levels based on this established mechanism that fosters indirect reciprocity. Factors that affect the effectiveness of hybrid scoring negatively and positively are group size and population size, respectively. The rate of return was shown to not matter.

There is one important component that the model we considered here externalises, namely the issue of how a hierarchy of scores translates into an analogous group formation hierarchy. These are questions future work should address. In particular, we need to address how mechanisms known to play such a role in the context of image scoring would translate into the informational setting considered here, and how such mechanisms may be designed. Little is known in this direction. Economic experiments \cite{rand_jtb12, rand_pnas14} might be particularly conductive to such research and guide future theoretical work to relevantly address these fundamental dilemmas of human cooperation.

\section*{Methods}

\subsection*{Simulation procedure}
The employed Monte Carlo simulation procedure \cite{binder_88} requires the iteration of the following three elementary steps. First, two randomly selected players $i$ and $j$ play one instance of the public goods game in their current group, thereby obtaining payoffs $\phi_{i}$ and $\phi_{j}$, respectively. Next, player $j$ adopts the strategy of player $i$ with the probability given by the Fermi function $W=1/\{1+\exp[(\phi_{j}-\phi_{i})/K]\}$, where $K=0.1$ quantifies the uncertainty by strategy adoptions \cite{szabo_pr07}. Each full Monte Carlo step gives a chance for every player to change its strategy once on average. The reported fractions of cooperators and defectors were determined in the stationary state.

\subsection*{Stability analysis}
At a given time $t$, there are four action-score pairs: cooperate-one ($c_i^t=s_i^t=1$), defect-one ($c_i^t=0$ and $s_i^t=1$), cooperate-zero ($c_i^t=1$ and $s_i^t=0$), and defect-zero ($c_i^t=s_i^t=0$). Suppose players are hardwired to play either action. However, scores change. Depending on the scoring mechanism, the action vector $c$ in period-$t$ implies a score (or a probability distribution of scores) for period $t+1$.  Assume that in some period all four action-score pairs are represented by positive population proportions, $p_{C1},p_{D1},p_{C0},p_{D0}$. Write $p_{C},p_{D}$ for the proportions of cooperators and defectors. Suppose the action proportions grow/ shrink, for action $a=\{C,D\}$, given by a replicator equation similar to that in the standard form: $\partial p_{a}/\partial t= p^t_{a1} *(\pi_{a1}-\overline{\pi})+p^t_{a0} *(\pi_{a0}-\overline{\pi})$, where $\pi_{as}$ is the expected payoff to action-score pair $as$ and $\overline{\pi}$ is the average population payoff.

\subsubsection*{Stability of unconditional defection under all scoring rules}
Suppose all players $j$ in period $t$, independent of their score, defect except for one player $i$ who plays $c_i^t=1$. W.l.o.g., suppose $s_i^t=1$.
No matter what $i$'s score, the payoff to $i$ representing $p_{C1}$ is $\pi_{C1}=r/s<1$, while the average payoff is $\overline{\pi}=1+\frac{s-1}{n-1}(r/s)>0$.
Hence, $\partial p_{C}/\partial t<0$ and therefore any such process is stable at $p_{C}=0$.

\subsubsection*{Stability of high cooperation levels under image scoring}
Suppose the four different strategies at time $t$ have mass of $p_{C1},p_{D1},p_{C0},p_{D0}$ respectively such that $p_{D1}=0$.
We shall now show that there exists a starting state with $p_C>(s-1)/s$ such that $\partial p_{C}/\partial t=0$.
Suppose that $p_{C1}=(n-s)/n$. Then  $\pi_{C1}= r$, $\pi_{C0}= (s - n*p_{D0})*(r/s)$ and $\pi_{D0}= 1+ (s - n*p_{D0})*(r/s)$.
$\partial p_{C}/\partial t=0$ if $\frac{n-s}{n}*r+  \frac{s - n*p_{D0}}{n} *(s - n*p_{D0})*(r/s)= p_{D0}*(1+ (s - n*p_{D0})*(r/s))$.
This yields $p_{DO}=1/(1-r+n*(r/s))$. Hence, a high cooperation level of $p_{C}=\frac{n*(r/s)-r}{1-r+n*(r/s)}$ is stable if $p_{C1}=(n-s)/n$, $p_{D1}=0$, $p_{C0}=(s/n)-1/(1-r+n*(r/s))$, and $p_{DO}=1/(1-r+n*(r/s))$. Notice that the temporary growth in $p_{C1}$ relative to $p_{C0}$ is compensated by the score dynamics which imply that exactly the proportion by which $p_{C0}$ will be replaced by the growth in $p_{C1}$.
Notice also that these proportions are robust to a small increase in $p_{D1}=0$ as the overproportional growth of $D1$ will result in scores of zero in the next period (all $D1$s will turn $D0$ in the next round), and then lead to a shrinking of $D0$.

\subsubsection*{Instability of high cooperation levels under group scoring}
At first sight, the state with $p_{C1}=(n-s)/n$, $p_{D1}=0$, $p_{C0}=(s/n)-1/(1-r+n*(r/s))$, and $p_{DO}=1/(1-r+n*(r/s))$ could be stable. Indeed, provided that $p_{D1}=0$ this state is stable against growth of proportions $p_{C1}$, $p_{C0}$ and $p_{DO}$. However, any small increase in $p_{D1}$ destabilizes the whole system. This is because, given $p_{C1}$, $p_{C0}$ and $p_{DO}$, the non-zero proportion $p_{D1}$ not only grows overproportional, but also -- and this is the key difference with image scoring-- is likely to keep a score of one, and will therefore also not shrink in the next period. Hence, the growth of $C1$ slows down while $D1$ keeps growing. There can be no stable state with a positive proportion $p_{D1}$ as both $p_{D1}$ and $p_{D0}$ would continue to grow faster than their cooperation counterparts.

\begin{acknowledgments}
This research was supported by the European Commission through the ERC Advanced Investigator Grant `Momentum' (Grant 324247), by the Deanship of Scientific Research, King Abdulaziz University (Grant 76-130-35-HiCi), by the Slovenian Research Agency (Grant P5-0027), and by the Hungarian National Research Fund (Grant K-101490).
\end{acknowledgments}

\end{document}